\documentclass[sigconf]{acmart}

\usepackage{bm}
\usepackage{booktabs}
\usepackage{url}
\usepackage{hyperref}
\usepackage{multirow}
\usepackage[ruled,vlined,boxed,linesnumbered]{algorithm2e}
\usepackage{caption}
\usepackage{graphicx}
\usepackage{subcaption}
\usepackage{enumitem}
\usepackage{balance}

\AtBeginDocument{%
  \providecommand\BibTeX{{%
    \normalfont B\kern-0.5em{\scshape i\kern-0.25em b}\kern-0.8em\TeX}}}

\copyrightyear{2022}
\acmYear{2022}
\setcopyright{acmcopyright}
\acmConference[SIGIR '22]{Proceedings of the 45th International ACM SIGIR Conference on Research and Development in Information Retrieval}{July 11--15, 2022}{Madrid, Spain} \acmBooktitle{Proceedings of the 45th International ACM SIGIR Conference on Research and Development in Information Retrieval (SIGIR '22), July 11--15, 2022, Madrid, Spain}
\acmPrice{15.00}
\acmDOI{10.1145/3477495.3531988}
\acmISBN{978-1-4503-8732-3/22/07}

\begin{document}

\fancyhead{}

\title{HIEN: Hierarchical Intention Embedding Network for Click-Through Rate Prediction}

% \author{Anonymous}
\author{Zuowu Zheng}
\authornote{Z. Zheng, X. Gao, and G. Chen are with the MoE Key Lab of Artificial Intelligence, Department of Computer Science and Engineering, Shanghai Jiao Tong University.}
\affiliation{%
  \institution{Shanghai Jiao Tong University}
  \city{Shanghai}
  \country{China}}
\email{waydrow@sjtu.edu.cn}
 
\author{Changwang Zhang}
\affiliation{%
  \institution{Tencent Inc.}
  \city{Shenzhen}
  \country{China}}
% \email{changwzhang@tencent.com}
\email{changwangzhang@foxmail.com}

\author{Xiaofeng Gao}
\authornote{This work was supported by the National Key R\&D Program of China [2020YFB1707903]; the National Natural Science Foundation of China [61872238, 61972254]; Shanghai Municipal Science and Technology Major Project [2021SHZDZX0102]; the CCF-Tencent Open Fund [RAGR20200105]; and the Tencent Marketing Solution Rhino-Bird Focused Research Program [FR202001]. X. Gao is the Corresponding author. Z. Zheng is also supported by 2021 Tencent Rhino-Bird Research Elite Training Program.}
\authornotemark[1]
\affiliation{%
  \institution{Shanghai Jiao Tong University}
  \city{Shanghai}
  \country{China}}
\email{gao-xf@cs.sjtu.edu.cn}

\author{Guihai Chen}
\authornotemark[1]
\affiliation{%
  \institution{Shanghai Jiao Tong University}
  \city{Shanghai}
  \country{China}}
\email{gchen@cs.sjtu.edu.cn}



\begin{abstract}
Click-through rate (CTR) prediction plays an important role in online advertising and recommendation systems, which aims at estimating the probability of a user clicking on a specific item. Feature interaction modeling and user interest modeling methods are two popular domains in CTR prediction, and they have been studied extensively in recent years. However, these methods still suffer from two limitations. First, traditional methods regard item attributes as ID features, while neglecting structure information and relation dependencies among attributes. Second, when mining user interests from user-item interactions, current models ignore user intents and item intents for different attributes, which lacks interpretability.
Based on this observation, in this paper, we propose a novel approach Hierarchical Intention Embedding Network (HIEN), which considers dependencies of attributes based on bottom-up tree aggregation in the constructed attribute graph. HIEN also captures user intents for different item attributes as well as item intents based on our proposed hierarchical attention mechanism.
Extensive experiments on both public and production datasets show that the proposed model significantly outperforms the state-of-the-art methods. In addition, HIEN can be applied as an input module to state-of-the-art CTR prediction methods, bringing further performance lift for these existing models that might already be intensively used in real systems.

\end{abstract}




\maketitle

\section{Introduction}
% background

Click-through rate (CTR) prediction is one of the most fundamental tasks for online advertising systems, and it has attracted much attention from both industrial and academical communities~\cite{chapelle2014simple,mcmahan2013ad,richardson2007predicting}. Most of the existing works in this field can be classified into two domains: feature interaction modeling and user interest modeling. In general, these methods follow a common paradigm, i.e., Embedding \& Multi-Layer Perceptron (MLP) learning. Raw sparse input features are first projected into dense embedding vectors, and then simply concatenated together to feed into deep neural networks (DNN) or other carefully designed neural networks to learn high-order feature interactions or user interests. In feature interaction modeling, Factorization Machines (FM) based methods are very popular, such as FM~\cite{rendle2010factorization}, Field-weighted factorization machine (FwFM)~\cite{pan2018field}, and DeepFM~\cite{guo2017deepfm}. The methods in user interest modeling focus on mining user interests from user historical behaviors including click, like, comment, etc., such as Deep Interest Network (DIN)~\cite{zhou2018deep}, Deep Interest Evolution Network (DIEN)~\cite{zhou2019deep}, and Deep Session Interest Network (DSIN)~\cite{feng2019deep}.

\begin{figure}[h]
	\centering
	\includegraphics[width=\linewidth,angle=0]{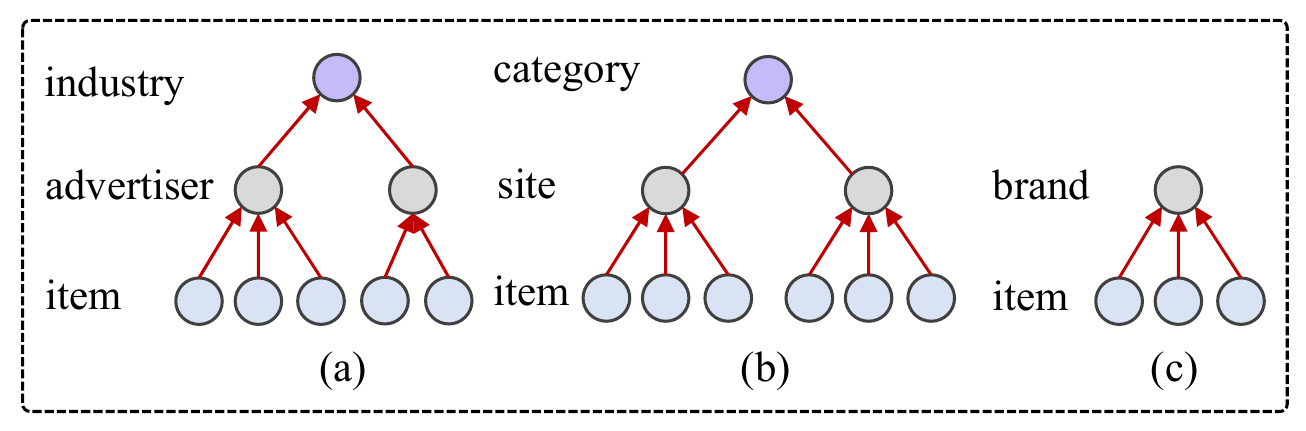}
	\caption{Illustration of relations between item attributes. Blue circles represent items, while the grey and purple circles representing attributes of item. The red arrow indicates the relations among attributes and items.}
	\label{fig:feature_tree}
\end{figure}

However, these models still suffer from the following challenges, which limit the performance improvement.
\begin{itemize}
    \item First, in online advertising systems, an item usually contains multiple attributes, e.g., \textit{item\_id}, \textit{category\_id}, \textit{advertiser\_id}, etc. Traditional methods converted these ID attributes to one-hot encoding vectors, and then embedded to dense real-value vectors for the following feature interactions. However, there are relations and dependencies between attributes of an item, which are ignored in existing methods. As shown in Fig.~\ref{fig:feature_tree}(a), an item belongs to an advertiser, an advertiser belongs to an industry. For example, a mobile-game (industry) company (advertiser) launched a new game (item), which contains the above relations. That is to say, using a single ID feature to represent an attribute is inadequate, the structural information among which should also be considered.
    \item Second, current user interest models focus on mining interests through interactions between user and item, while neglecting user intents and item intents for different attributes. E.g., a user may click a new game ad due to its category and release time. In contrast, an item may be clicked by a user due to his or her age and occupation.  Existing user interest models fail to reveal these intents for different attributes, which lacks interpretability.
\end{itemize}

Currently, graph neural network (GNN) achieves significant success in recommendation system, which can model structure information and relations among nodes in constructed graph, such as LightGCN~\cite{DBLP:conf/sigir/0001DWLZ020}, Neural Graph Collaborative Filtering (NGCF)~\cite{DBLP:conf/sigir/Wang0WFC19}, and Knowledge Graph Attention Network (KGAT)~\cite{DBLP:conf/kdd/Wang00LC19}. It is intuitive to introduce GNN to consider attribute dependencies of item. However, most of existing GNN based works perform graph convolution to aggregate information from neighbor nodes, ignoring different characteristics of different attributes.

In this paper, we propose a novel approach Hierarchical Intention Embedding Network (HIEN), which is designed with two considerations to address the above two challenges in existing methods. On the one hand, we construct two types of graphs, i.e., item-attribute graph and user-attribute graph. Some of the attributes and items form a tree structure, as shown in Fig.~\ref{fig:feature_tree}. In this type of graph, we perform attribute 
tree convolution and aggregate child nodes from the bottom-up, which is able to capture structure information of item attributes. In this way, we can refine attribute representations. On the other hand, when learning user and item representations, we consider user intents for different item attributes, as well as item intents for user attributes. Specifically, we propose hierarchical attention mechanism to consider attribute hierarchy. The information of refined attribute embedding are propagated to user and item embedding with hierarchical attentive weights. Furthermore, HIEN serves as an embedding learning framework, which works compatibly with the existing deep CTR models.

The main contributions of this paper are summarized as follows:
\begin{itemize}
    \item We propose a novel approach Hierarchical Intention Embedding Network named HIEN for click-through rate prediction, enhancing the feature embeddings with structural information between attributes. To the best of our knowledge, this is the first deep CTR model that considers attribute relations and dependencies.
    \item We propose dual intentions including user and item intents to learn user and item representations explicitly. Specifically, a hierarchical attention mechanism is carefully designed to capture inherent importance and dynamic effects from different attributes.
    \item We conduct extensive experiments on both public and production datasets in real-world online advertising system with the state-of-the-art methods. Evaluation results verify the effectiveness of the proposed method in Embedding \& MLP models for CTR prediction.
\end{itemize}

The rest of the paper is organized as follows. Section 2 provides the preliminaries of existing deep CTR models. In Section 3, we introduce the proposed HIEN. Experimental settings and evaluation results are presented in Section 4. Finally, Section 5 and Section 6 discusses the related works and concludes the paper, respectively.

\section{Preliminaries}
In this section, we revisit the preliminaries of CTR prediction task, including problem formulation and a general Embedding \& MLP paradigm named Base Model.
\subsection{Problem Formulation}
A CTR prediction model aims at predicting the probability that a user $u$ clicks an item $v$, which mostly takes three-tuple as input features:
\begin{equation}
    p\mbox{CTR} = f(\mbox{user}, \mbox{item}, \mbox{context})\nonumber
\end{equation}
where user fields group contains user profiles and user behaviors, item and context group contains features from the item and context side, respectively.
We denote user set $U=\{u_1, u_2, ..., u_M\}$ and item set $V=\{v_1, v_2, ..., v_N\}$, where $M$ and $N$ are the number of users and items, respectively.
For each sample, there are a set of $J$ fields of user attributes $A_u=\{a_u^1, a_u^2,..., a_u^J\}$, a set of $K$ fields of item attributes $A_v=\{a_v^1, a_v^2, ..., a_v^K\}$, and a set of $P$ fields of context features $C=\{c^1, c^2, ..., c^P\}$ including click time, location, system information, etc. User-item interactions can be denoted as a matrix $Y\in \mathcal{R}^{M\times N}$, where $y_{uv}=1$ denotes user $u$ clicks item $v$, otherwise $y_{uv}=0$. In addition, each user $u$ has a historical click behavior sequence $B_u=\{b_1, b_2, ..., b_{L_u}\}$, where $b_i\in V$ and $L_u$ denotes the behavior length of user $u$.
Then one sample can be represented as
\begin{equation}
    x=\{u, v, A_u, A_v, B_u, C\}
\end{equation}
where $u$ and $v$ denote \textit{user\_id} and \textit{item\_id} in this paper, respectively.

\begin{figure*}[ht]
    \centering
    \includegraphics[width=.85\linewidth]{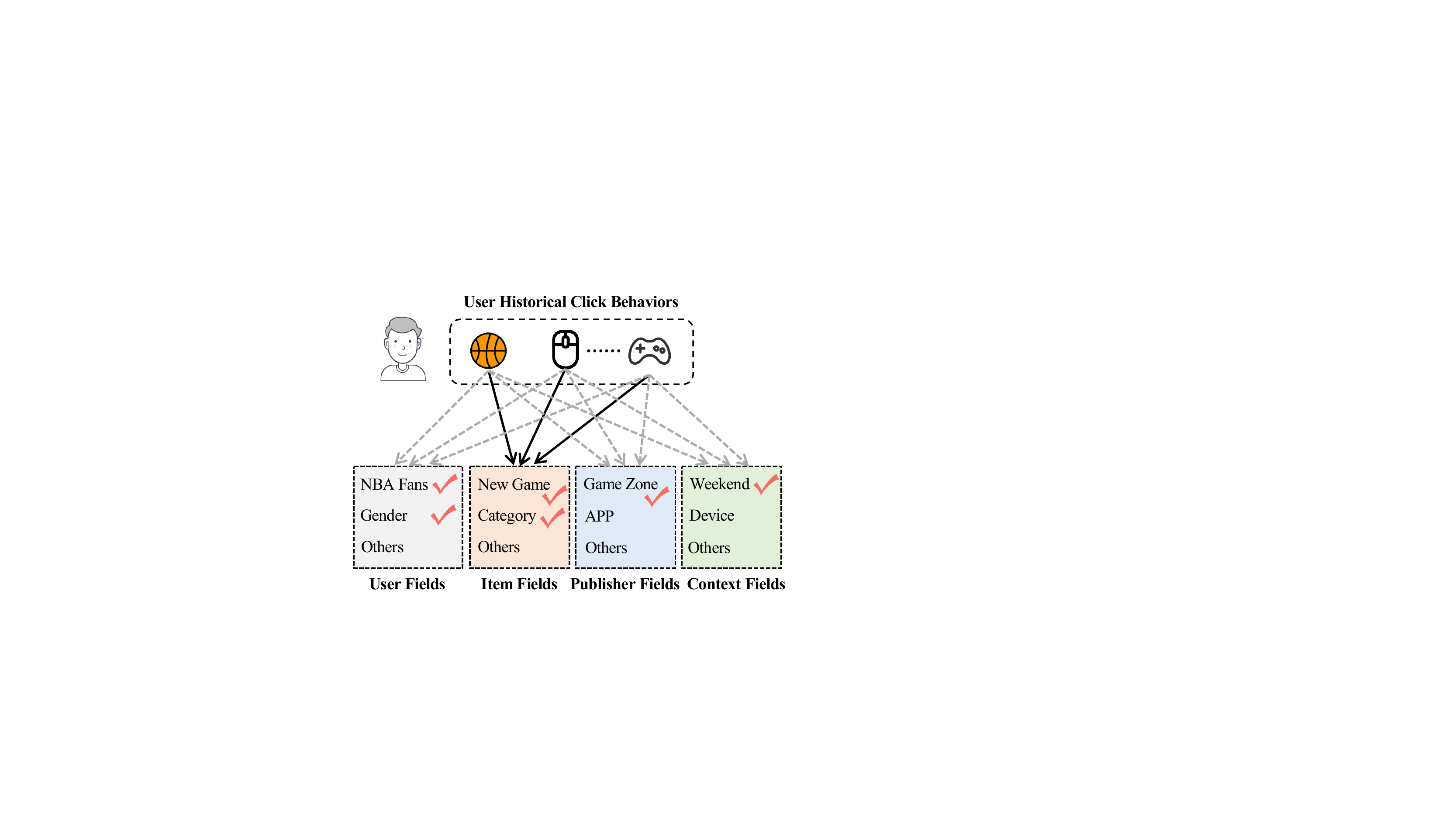}
    \caption{Architecture of the proposed framework, HIEN. Attribute tree aggregation considers structural information and relation dependencies between attributes, the embedding vectors of which are refined through bottom-up attribute tree aggregation. In order to model user and item intents for different attributes, the intention modeling module is designed based on hierarchical attention and graph neural networks. The refined user embeddings, item embeddings, and attribute embeddings are fed into deep CTR network to produce final score.}
    \label{fig:framework}
\end{figure*}

\subsection{Base Model}
\label{sec:base_model}
Existing CTR prediction models follow a general Embedding \& MLP paradigm.
\subsubsection{Feature Embedding}
We first transform it into high-dimensional sparse binary features via encoding. For example, the encoding vector of item id is formulated as $v\in R^{D}$. $D$ is the dimension of $\bm{v}$, i.e., there are $D$ unique possible values in $v$. $\bm{v}[i]\in\{0,1\}$ is the $i$-th value of $\bm{v}$. $\sum_{i=1}^D\bm{v}[i]=1$, which is a one-hot encoding, i.e, $[0,0,1,...,0]$. 
Then we use embedding technique to transform high dimensional binary vectors into low dimensional dense representations. Let $\bm{W}\in\mathcal{R}^{D\times K}$ be the embedding dictionaries of $\bm{v}$. For one-hot encoding vector $\bm{v}$, the embedding representation of $\bm{v}$ is $\bm{e}_v\in\mathcal{R}^K$. For multi-hot encoding vector like user behaviors $B_u$, the embedding representation is a set of vectors. The representation of other features are similar, which is omitted for simplicity. The whole input feature embeddings can be represented as
\begin{equation}
    \bm{E} = \{\bm{e}_u, \bm{e}_v, \bm{e}_{A_u}, \bm{e}_{A_v}, \bm{e}_{B_u}, \bm{e}_C\}
\end{equation}

\subsubsection{Feature Interaction}
Existing CTR models commonly design complicated network structures to learn high-order feature interactions. DeepFM~\cite{guo2017deepfm} is a widely used deep CTR models, which combines the power of factorization machines for recommendation and deep learning for feature learning in a new neural network architecture. We take DeepFM as our Base Model in this paper.
\begin{equation}
    \hat{y} = \mbox{sigmoid}(y_{FM} + y_{DNN})
\end{equation}
where $\hat{y}$ is the predicted CTR score, $y_{FM}$ is the output of FM component, and $y_{DNN}$ is the output of DNN.
\begin{equation}
    y_{FM} = \sum_{i=1}^{N_f}x_iw_i + \sum_{i=1}^{N_f}\sum_{j=i+1}^{N_f}x_ix_j\langle\bm{e_i},\bm{e_j}\rangle
\end{equation}
where $N_f$ is the number of unique features. $y_{FM}$ is able to learn 1-order features and 2-order feature interactions with dot product of two feature embeddings. Denote the output of feature embedding layer as
\begin{equation}
    a^{(0)} = \bm{E} = [\bm{e}_1, \bm{e}_2, ..., \bm{e}_m],
\end{equation}
where $\bm{e}_i$ is the embedding of $i$-th field and $m$ is the number of all fields. Then we feed $a^{(0)}$ into DNN, and the forward process is
\begin{equation}
    a^{(l+1)} = \sigma(W^{(l)}a^{(l)}+b^{(l)}),
\end{equation}
where $l$ is the depth of MLP layer and $\sigma$ is the activation function, $W^{(l)}$ and $b^{(l)}$ is the weight and bias of $l$-th layer respectively.
\begin{equation}
    y_{DNN} = \sigma(W^{(L)}a^{(L-1)}+b^{(L)})
\end{equation}
where $L$ is the number of hidden layers of DNN.

\subsubsection{Model Training}
We minimize the following cross-entropy loss in model training:
\begin{equation}
    L(\Theta) = -\frac{1}{|D|}\sum_{(\bm{x},y)\in\mathcal{D}}(y\log \hat{y}+(1-y)\log (1-\hat{y})) + \lambda\left\|\Theta\right\|_2
\end{equation}
where $\mathcal{D}$ is the training dataset, $\Theta$ includes all trainable parameters, $x$ is the input of the model, $y\in\{0,1\}$ is the label that indicates whether the user clicked the item, $\hat{y}$ is the predicted probability that the user clicks the item. $L_2$ regularization weighted by $\lambda$ is adopted to $\Theta$ to prevent overfitting.

\section{Hierarchical Intention Embedding Network}

Most of the existing works focus on feature interaction layers to learn high-order and more expressive feature interactions, while neglecting optimization of feature embedding layers. We aim to learn structural information of attributes and user/item intentions, which refines the original feature embeddings. In this section, we propose Hierarchical Intention Embedding Network (HIEN), which contains three modules: graph construction, attribute tree aggregation, and hierarchical intention modeling. We elaborate on each of these three modules in detail. Fig.~\ref{fig:framework} gives an illustration of the proposed HIEN framework.

\subsection{Graph Construction}
\subsubsection{Attribute Graph}
Since multiple users or items may have same attributes, we can connect them and construct a graph. We first construct item-attribute graph $G_v=(V\cup A_v, \mathcal{E}_v)$, where $\mathcal{E}_v$ is the set of edges. An edge $e_v=(v,a_v)$ indicates that item $v$ has attribute $a_v$. As shown in Fig.~\ref{fig:feature_tree}, there are relations and dependencies between attributes of item. Correspondingly, we separate attributes and items into multiple tree-like feature structures (i.e., item-attribute tree $T_v$) where each tree contains all connected attributes and items. Then item-attribute graph $G_v$ can be regarded as a set of item-attribute tree, i.e., $G_v^t=\{T_v^1, T_v^2, ..., T_v^n\}$. An edge $e_t=(t_i, t_j)$ in $T_v$ denotes there is a relation, where $t_i\in V\cup A_v$ and $t_j\in A_v$ are nodes in tree $T_v$. In each item-attribute tree, items are leaf nodes and attributes are parent nodes. Similarly, we can construct user-attribute graph $G_u=\{U\cup A_u, \mathcal{E}_u\}$ and user-attribute tree $G_u^t$. Note that there is no obvious hierarchy in user attributes. However, the proposed general method can also be applied in the user-attribute graph.

\subsubsection{User-Item Bipartite Graph}
In online advertising system scenario, we typically have user-item interactions $Y$ (e.g., clicks and likes), which can be utilized to capture user interests. Here we define user-item bipartite graph $G_{uv}=(U\cup V, \mathcal{E}_{uv})$ from user-item interaction matrix $Y$, of which an edge $e_{uv}=(u,v)$ if $y_{uv}=1$.

\subsubsection{Task Description}
We now formulate the prediction task to be addressed in this paper:
\begin{itemize}
    \item Input: attribute tree that includes the item-attribute tree set $G_v^t$ and the uesr-attribute tree set $G_u^t$, user-item bipartite graph $G_{uv}$.
    \item Output: a prediction function that predicts the probability $\hat{y}_{uv}$ that user $u$ would click item $v$. 
\end{itemize}

\subsection{Attribute Tree Aggregation}
\label{sec:attribute_tree}
When constructing attribute graph $G_u$ and $G_v$, an intuitive way to capture structure information and enrich nodes representation is to perform graph convolution. It should be noted that the characteristics of different attributes vary greatly. For example, \textit{advertiser\_id} and \textit{item\_price} from item attributes set have different semantics and distributions, and aggregating them in graph learning may introduce noise. However, most of the existing GNN based methods perform graph convolution to aggregate information from neighbor nodes, while neglecting different characteristics between different attributes~\cite{DBLP:conf/sigir/0001DWLZ020,DBLP:conf/sigir/Wang0WFC19}. In this section, we propose a bottom-up aggregation strategy to consider structure feature information, which maintains inherent characteristics of different attributes. Specifically, we perform aggregation on attribute tree $G_v^t$ and $G_u^t$ individually.

For each attribute tree $T$ in $G_v^t$ and $G_u^t$, we need to learn representation of node $\bm{e}_h$. We propose to learn $\bm{e}_h$ through aggregating its child nodes $\bm{e}_{C_h}$ from the bottom-up manner. Formally, $\bm{e}_h = g(\bm{e}_h, \bm{e}_{C_h})$. We implement $g(\cdot)$ based on several state-of-the-art GCN models for the attribute tree aggregation.

\begin{itemize}
    \item GCN Aggregator. GCN~\cite{DBLP:conf/iclr/KipfW17} sums up the representation of the central node and its directly connected nodes and then applies a non-linear transformation, which can be formulated as follows:
    \begin{equation}
        g_{GCN}(\bm{e}_h, \bm{e}_{C_h}) = \sigma(\bm{W}(\bm{e}_h + \bm{e}_{C_h}))
        \label{eqn:gcn}
    \end{equation}
    where $\sigma$ is a non-linear activation function and $\bm{W}$ is the trainable weight matrix to distill useful information.
    \item NGCF Aggregator. NGCF~\cite{DBLP:conf/sigir/Wang0WFC19} considers feature interaction between the central node and its neighbor nodes. It first does element-wise product to calculate the interaction of central node and neighbors and then adds neighbors representation to central node representation.
    \begin{equation}
        g_{NGCF}(\bm{e}_h, \bm{e}_{C_h}) = \sigma(\bm{W}_1\bm{e}_h + \sum_{i\in C_h}(\bm{W}_1\bm{e}_i + \bm{W}_2(\bm{e}_h \odot \bm{e}_i)))
    \end{equation}
    where $\bm{W}_1$ and $\bm{W}_2$ are the trainable weight matrix and $\odot$ denotes the element-wise product.
    \item LightGCN Aggregator. LightGCN~\cite{DBLP:conf/sigir/0001DWLZ020} argues that feature transformation and non-linear activation contribute little to the model performance, which adds to the difficulty of training and even degrades the recommendation performance.
    \begin{equation}
        g_{LightGCN}(\bm{e}_h, \bm{e}_{C_h}) = \sum_{i\in C_h} \bm{e}_i
    \end{equation}
    \item Concat \& Product Aggregator (CP-Agg). We carefully design the new aggregator that considers two types of interactions between the central node and its child nodes.
    \begin{equation}
    \begin{split}
        &g_{Concat\&Product}(\bm{e}_h, \bm{e}_{C_h}) = f(\bm{e}_h\odot\bm{e}_{C_h}, \bm{e}_h\oplus\bm{e}_{C_h}) \\
        &= \sigma(\bm{W}_1\bm{e}_h + \sum_{i\in C_h}(\bm{W}_1\bm{e}_i + \bm{W}_2(\bm{e}_h \odot \bm{e}_i) + \bm{W}_3(\bm{e}_h \oplus \bm{e}_i)))
    \end{split}
    \label{eqn:cp-agg}
    \end{equation}
    where $\oplus$ denotes concatenation and $\bm{W}_1$, $\bm{W}_2$, $\bm{W}_3$ are all trainable weight matrices that transform the interaction result into a uniformed vector of the same K dimension.
\end{itemize}
We empirically compare these aggregators in Sec.~\ref{sec:hien_study}.
Through attribute tree aggregation, the representations of item attributes and user attributes are refined, which contains structure relations and dependencies between attribute. The input feature embeddings in data can be refined as:
\begin{equation}
    \bm{E}^\prime = \{\bm{e}_u, \bm{e_v}, \bm{e}_{A_u}^\prime, \bm{e}_{A_v}^\prime, \bm{e}_{B_u}, \bm{e}_C\}
\end{equation}

\subsection{Hierarchical Intention Modeling}
\begin{figure}
    \centering
    \includegraphics[width=\linewidth]{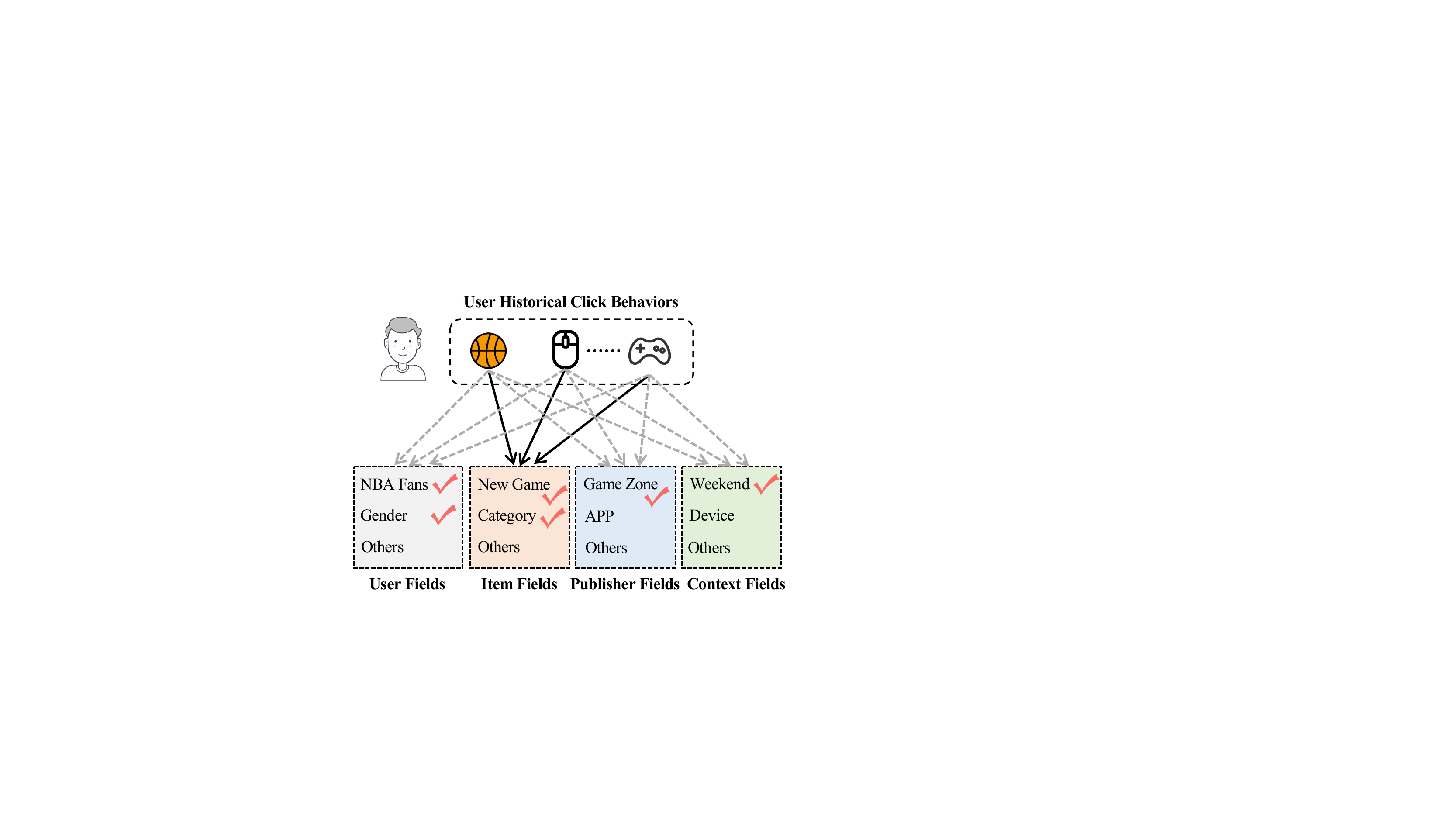}
    \caption{Toy example of hierarchical intention modeling. The nodes denote different attributes. The red arrows represent relations and dependencies between attributes, where $\alpha_i$ and $\beta_i$ measure the inherent importance of attributes. The black/grey arrows represent user intent and item intent for the corresponding attributes, where $a_i$ and $b_i$ are dynamic scores of different attributes when modeling intention.}
    \label{fig:intent_fig}
\end{figure}
After we refine attribute representation, we aim at learning user and item embedding representations and capturing user interests. Existing user interests methods can be divided into two categories: user behavior modeling and GNN based methods. The former one like DIN~\cite{zhou2018deep} and DIEN~\cite{zhou2019deep} focuses on modeling user historical behaviors through sequential modeling, e.g., RNN and Transformer, which considers the relations between the target item and user behaviors. The latter one exploits the user-item bipartite graph by propagating embeddings on it, which leads to the expressive modeling of high-order connectivity in the user-item graph. However, most of these methods neglect user intents and item intents for different attributes. For example, a user may click a new game ad due to its category and release time. In contrast, an item may be clicked by a user due to his or her age and occupation. Existing user interest models fail to reveal these intents, and they lack interpretability.

In this section, we propose to consider dual intents from user and item sides respectively based on user-item graph learning, which can enrich user and item representations during the learning process. Given user-item bipartite graph $G_{uv}$, we use user-item edge $\mathcal{E}_{uv}$ to learn their embeddings. In order to consider user intents for different item attributes, we assign an attentive weight for each item attribute. In addition, we also consider the inherent effects of hierarchical attributes. As shown in Fig.~\ref{fig:intent_fig}, the item embedding $\bm{e}_v^{(1)}$
% \begin{equation}
%     \bm{e}_v^{(1)} = \bm{e}_v^{(0)} + \sum_{a\in A_v}\alpha(u, a)\bm{e}_a,
% \end{equation}
\begin{equation}
    \bm{e}_v^{(1)} = a_0\alpha_0\bm{e}_x^0 + a_1\alpha_1\bm{e}_x^1 + a_2(\alpha_2\bm{e}_x^2 + a_4\alpha_4\bm{e}_x^4) + a_3\alpha_3\bm{e}_x^3,
    \label{eqn:user_intent_att}
\end{equation}
where $a_i$ is the attention score to differentiate the intent of user $u$ for item attribute $\bm{e}_x^i$. $\alpha_i$ denotes the inherent importance of item attribute and is a learnable parameter during the training process. Please note that when assigning weights for $\bm{e}_x^2$ and $\bm{e}_x^4$, we have considered their hierarchical structure since $\bm{e}_x^4$ (\textit{industry\_id}) is the parent node of $\bm{e}_x^2$ (\textit{advertiser\_id}). In Eqn.~\eqref{eqn:user_intent_att} we introduce a concise and elegant formula $a_2(\alpha_2\bm{e}_x^2 + a_4\alpha_4\bm{e}_x^4)$ to inject the hierarchical relations between attributes into the attention mechanism.

$a_i$ is implemented based on inner product as
\begin{equation}
    a_i(u,x_i) = \frac{\exp(\bm{e}_u^{(0)}\bm{e}_x^i)}{\sum_{x^\prime\in A_v}\exp(\bm{e}_u^{(0)}\bm{e}_{x^\prime})},
\end{equation}
where $\bm{e}_u^{(0)}$ is the original ID embedding of user $u$ to make the importance score personalized. $a(u,x)$ gives us an intuitive way to explain which item attributes are more important for user $u$ to decide whether to click item $v$.

Similarly, we have initial user representation integrated with item intents for user attributes $\bm{e}_y$.
\begin{equation}
    \bm{e}_u^{(1)} = \sum_{i=0}b_i\beta_i\bm{e}_y^i,
\end{equation}
where $b_i$ reflects attention score of item $v$ for attribute $\bm{e}_y^i$ of user $u$. $\beta_i$ denotes inherent importance of user attribute $\bm{e}_y^i$ and is a trainable parameter. $b_i$ is calculated as:
\begin{equation}
    b_i(v,y_i) = \frac{\exp(\bm{e}_v^{(0)}\bm{e}_y^i)}{\sum_{y^\prime\in A_u}\exp(\bm{e}_v^{(0)}\bm{e}_{y^\prime})},
\end{equation}
where $\bm{e}_v^{(0)}$ is the original ID embedding of item $v$. Then we can model user-item interactions and generate user and item representations by
\begin{equation}
    \bm{e}_u^{(l)} = g(\bm{e}_u^{(l-1)}, \bm{e}_{N_u}^{(l-1)})
\end{equation}
\begin{equation}
    \bm{e}_v^{(l)} = g(\bm{e}_v^{(l-1)}, \bm{e}_{N_v}^{(l-1)})
\end{equation}
where $N_u$ and $N_v$ denotes neighbor nodes of user $u$ and item $v$ in user-item bipartite graph $G_{uv}$, respectively. $g(\cdot)$ is the aggregation function mentioned in Sec.~\ref{sec:attribute_tree}. It is worth noting that, here we use the user-item bipartite graph $G_{uv}$ rather than the tree structure to conduct the graph aggregation.

\subsection{Model Prediction}
After $L$ convolution layers, we can obtain the embedding representations of user $u$ and item $v$ at different layers and then sum them up as the final representations.
\begin{equation}
\begin{split}
    \bm{e}_u^\prime &= \bm{e}_u^{(0)} + \bm{e}_u^{(1)} + ... + \bm{e}_u^{(L)} \\
    \bm{e}_v^\prime &= \bm{e}_v^{(0)} + \bm{e}_v^{(1)} + ... + \bm{e}_v^{(L)}
\end{split}
\end{equation}

The whole input features can be refined as
\begin{equation}
    \bm{E}^\prime = \{\bm{e}_u^\prime, \bm{e}_v^\prime, \bm{e}_{A_u}^\prime, \bm{e}_{A_v}^\prime, \bm{e}_{B_u}^\prime, \bm{e}_C\}
\end{equation}
where intent-aware relations are encoded in the user and item representation, structural information between attributes are encoded in the attribute representation. Finally, $\bm{E}^\prime$ is fed into Base Model in Sec.~\ref{sec:base_model} for training and output the predicted CTR score.

\section{Experiment}
\begin{table*}[t]
    \centering
    \caption{Statistics of the datasets.}
    \setlength{\tabcolsep}{5mm}{
    \begin{tabular}{c|c|c|c|c|c|c}
    \toprule
    Datasets & \#Train Samples & \#Test Samples & \#Fields & \#Features & \#Items & Positive Ratio\\
    \midrule
    Alibaba & 5,544,213 & 660,694 & 16 & 1,657,981 & 512,431 & 5.147\%\\
    Tmall & 4,382,613 & 505,228 & 9 & 1,480,725 & 565,888 & 10.230\%\\
    Tencent & 5,240,498 & 873,604 & 17 & 1,248,065 & 347,206 & 12.998\%\\
    \bottomrule
    \end{tabular}}
    \label{tab:datasets}
\end{table*}
In this section, we evaluate our proposed model on three real-world datasets: the public Alibaba Display Ad CTR dataset, the public Tmall dataset, and our Tencent CTR dataset in real advertising system. We aim to answer the following research questions:
\begin{itemize}
    \item \textbf{RQ1}: How does our method HIEN perform, comparing to the state-of-the-art CTR models?
    \item \textbf{RQ2}: What is the impact of the designs (e.g., the graph aggregators, different components, layer numbers) on the performance of HIEN?
    \item \textbf{RQ3}: The proposed HIEN can be used as an input module in other CTR models. Dose the refined input embedding vectors by HIEN bring performance lift?
    \item \textbf{RQ4}: Can HIEN provide insights on intent modeling and give an intuitive impression of explainability?
\end{itemize}

\subsection{Datasets Description}
We use both public and production datasets to evaluate the effectiveness of the proposed model. The statistics of the datasets is shown in Tab.~\ref{tab:datasets}.

\begin{itemize}
    \item \textbf{Alibaba Dataset\footnote{\url{https://tianchi.aliyun.com/dataset/dataDetail?dataId=56}}.} Alibaba advertising dataset~\cite{feng2019deep} is a public dataset released by Alibaba, which is one of the largest online commercial advertising platforms in the world. It randomly sampled 1,140,000 users from the website of Taobao for 8 days of ad display and click logs (26 million records) to form the original dataset. We used 7 days’s samples as the training dataset (2017-05-06 to 2017-05-12), and the next day’s samples as the test dataset (2017-05-13). 
    % The goal is to predict whether a user will click a target item with multiple features. When extracting user behaviors features, we keep users' most recent 50 behaviors from logs in each dataset. Please note that we only extract the corresponding user click behaviors whose click time was before the target item to prevent information leakage. % There are 16 fields in Alibaba dataset, which can be categorized into three groups: user side fields such as user\_id, cms\_segid, cms\_groupid, gender, age\_level, consumption, shopping\_level, occupation, and city\_level; item side fields such as adgroup\_id, cate\_id, campaign\_id, brand, customer\_id, and price; publisher side fields such as scenario\_id. Among them, price of item side is a dense feature. We use the 15 fields to interact with user behaviors.
    \item \textbf{Tmall Dataset\footnote{\url{https://tianchi.aliyun.com/dataset/dataDetail?dataId=42}}.} Tmall dataset~\cite{DBLP:conf/kdd/GuoSTGZLTH21} is a public dataset provided by Tmall.com in IJCAI-15 contest, which contains anonymized users' shopping logs in the past 6 months before and on the "Double 11" day. The user profile is described by user ID, age range and gender. The item attributes include item ID, category, brand, and merchant ID. The context features are timestamp and action type.
    \item \textbf{Tencent Dataset.} This dataset is extracted from the click log of Tencent online advertising platforms. It was collected through sampling user click logs during one week in 2021, which contains 6 millions samples. Logs of the first six days are used for training and logs of the last day are used for testing. 
    % The data preprocessing strategy is the same as that of Alibaba dataset.
    % Logs from 2021-10-01 to 2021-10-06 are used for training and logs from 2021-10-07 are used for testing. The data preprocessing strategy is the same as that of Alibaba dataset.
    
    %This dataset was collected through sampling user click logs during one week, which contains 7 millions samples. Users' historical behahaviors include the click-through history of the past 6 months. Logs from 2021-04-23 to 2021-04-28 are used for training and logs from 2021-04-29 are used for testing. The data preprocessing strategy is the same as that of Alibaba dataset. 
    % Our production dataset consists of 15 fields, which includes three groups: user side fields such as user\_grade, user\_province, user\_age, and user\_gender; item side fields such as ad\_id, campaign\_id, creative\_id, advertiser\_id, ad\_product\_type, ctr\_size, ams\_first\_industry\_id, and ams\_second\_industry\_id; context side fields os\_type, device\_alias, and device\_brand.
\end{itemize}

It should be noted that most of public datasets lack enough details to test our model.
% , and Alibaba and Tmall datasets are, to the best of our knowledge, the only two public datasets that are applicable. 
We need feature semantics to capture structure relations between attributes, as well as user information to explore user/item intents. Most of existing public datasets are not applicable due to anonymous features (e.g., Criteo and Avazu), missing user information (e.g., Avazu), insufficient attributes (e.g., Amazon and MovieLens), and missing deep hierarchy of item attributes (e.g., Alipay). However, hierarchy and intention of attributes are very common but neglected information in commercial systems, which plays important roles in online advertising. We will verify that in the following experiments.

\subsection{Baseline Methods}
To verify the effectiveness of the proposed HIEN model, we compare it with four categories of CTR models: shallow models (LR~\cite{chapelle2014simple}, FM~\cite{rendle2010factorization}, FwFM~\cite{pan2018field}), deep models (DeepFM~\cite{guo2017deepfm}, AutoInt+~\cite{DBLP:conf/cikm/SongS0DX0T19}, PNN~\cite{qu2016product}), user interest models (DIN~\cite{zhou2018deep}, DIEN~\cite{zhou2019deep}, DSIN~\cite{feng2019deep}), and GNN-based models (GIN~\cite{DBLP:conf/sigir/LiCWRZZ19}, Fi-GNN~\cite{DBLP:conf/cikm/LiCWZW19}, DG-ENN~\cite{DBLP:conf/kdd/GuoSTGZLTH21}).

\subsection{Experimental Settings}
All methods are implemented in Tensorflow 1.4 with Python 3.5, which are trained from scratch without any pre-training on NVIDIA TESLA M40 GPU with 24G memory. For baseline methods, we refer to the hyper-parameter settings in their original papers but also finetune them on our datasets. The embedding dimension $K$ is 128 for all features. The last three MLP layers in our BaseModel (i.e. DeepFM) have dimensions of 200, 800, and 1, employing the activation functions of PReLU, PReLU, and Sigmoid respectively. We use Adagrad~\cite{duchi2011adaptive} as the optimizer, with a learning rate of 0.001. The batch size is 4,096 and 16,384 for training and test dataset, respectively. We run each experiments 5 times and report the average results.
% For user interest methods, such as DIN, DIEN, and DSIN, the dimension of two-layer MLP in the local activation unit is 200 and 1, with activation function dice~\cite{zhou2018deep}. For DSIN, we divide user behavior sequences into 5 sessions. The maximum user behavior length of each session is 10. For GNN based methods, we tune the GCN layer size for graph models in the range of $\{1,2,3,4\}$.

We use Area Under ROC Curve (AUC) as the evaluation metric, which is a widely used metric in CTR prediction tasks. It is defined as follows:
\begin{equation}
    \mbox{AUC} = \frac{1}{N^+N^-}\sum_{x^+\in D^+}\sum_{x^-\in D^-}(\mathcal{I}(f(x^+)>f(x^-)))
\end{equation}
where $D^+$ (resp. $D^-$) is the collection of all positive (resp. negative) samples. $f(x)$ is the predicted value with the input sample $x$ and $\mathcal{I}(\cdot)$ is the indicator function.

\subsection{Performance Comparison (RQ1)}
\begin{table*}[ht]
    \centering
    \caption{Experiment results of our model and competitors on the public Alibaba and Tmall datasets. The bold value marks the best one in each column, while the underlined value corresponds to the best one among all the baselines.}
    \setlength{\tabcolsep}{2.5mm}{
    \begin{tabular}{c|c|c|c|c|c|c}
    \toprule
    \multirow{2}{*}{Model} & \multicolumn{3}{c|}{Alibaba} & \multicolumn{3}{c}{Tmall}\\
    & Loss (mean$\pm$std) & AUC (mean$\pm$std) & AUC Impv. & Loss (mean$\pm$std) & AUC (mean$\pm$std) & AUC Impv.\\
    \midrule
    LR & 0.2638$\pm$0.00016 & 0.6157$\pm$0.00005 & - & 0.1985$\pm$0.00034 & 0.8724$\pm$0.00028 & -\\
    FM & 0.2571$\pm$0.00009 & 0.6283$\pm$0.00018 & 2.046\% & 0.1826$\pm$0.00004 & 0.8943$\pm$0.00011 & 2.510\% \\
    FwFM & 0.2526$\pm$0.00023 & 0.6427$\pm$0.00032 & 4.385\% & 0.1791$\pm$0.00057 & 0.9075$\pm$0.00030 & 4.023\% \\
    \midrule
    DeepFM & 0.2447$\pm$0.00010 & 0.6594$\pm$0.00035 & 7.098\% & 0.1772$\pm$0.00008 & 0.9184$\pm$0.00052 & 5.273\% \\
    AutoInt+ & 0.2506$\pm$0.00008 & 0.6508$\pm$0.00005 & 5.701\% & 0.1789$\pm$0.00026 & 0.9126$\pm$0.00044 & 4.608\% \\
    PNN & 0.2471$\pm$0.00037 & 0.6584$\pm$0.00076 & 6.935\% & 0.1783$\pm$0.00009 & 0.9160$\pm$0.00017 & 4.998\% \\
    \midrule
    DIN & 0.2364$\pm$0.00034 & 0.6523$\pm$0.00042 & 5.944\% & 0.1785$\pm$0.00046 & 0.9155$\pm$0.00019 & 4.940\% \\
    DIEN & 0.2321$\pm$0.00019 & 0.6610$\pm$0.00013 & 7.357\% & 0.1746$\pm$0.00006 & 0.9248$\pm$0.00039 & 6.006\% \\
    DSIN & 0.2308$\pm$0.00070 & 0.6708$\pm$0.00047 & 8.949\% & 0.1684$\pm$0.00021 & 0.9313$\pm$0.00063 & 6.751\% \\
    \midrule
    GIN & 0.2382$\pm$0.00007 & 0.6511$\pm$0.00068 & 5.750\% & 0.1779$\pm$0.00050 & 0.9150$\pm$0.00062 & 4.883\% \\
    Fi-GNN & 0.2414$\pm$0.00062 & 0.6502$\pm$0.00046 & 5.603\% & 0.1780$\pm$0.00032 & 0.9134$\pm$0.00009 & 4.700\% \\
    DG-ENN & \underline{0.2265}$\pm$0.00083 & \underline{0.6740}$\pm$0.00022 & 9.469\% & \underline{0.1635}$\pm$0.00017 & \underline{0.9396}$\pm$0.00044 & 7.703\% \\
    \midrule
    HIEN (ours) & \textbf{0.2156$\pm$0.00042} & \textbf{0.6798$\pm$0.00027} & \textbf{10.411\%} & \textbf{0.1601$\pm$0.00018} & \textbf{0.9467$\pm$0.00035} & \textbf{8.517\%} \\
    \bottomrule
    \end{tabular}}
    \label{tab:exp_public}
\end{table*}

\begin{table}[ht]
    \centering
    \caption{Experiment results of our model and competitors on the production dataset.}
    \begin{tabular}{c|c|c|c}
    \toprule
    Model & Loss (mean$\pm$std) & AUC (mean$\pm$std) & AUC Impv.\\
    \midrule
    LR & 0.3668$\pm$0.00068 & 0.7212$\pm$0.00057 & -\\
    FM & 0.3654$\pm$0.00071 & 0.7287$\pm$0.00052 & 1.040\% \\
    FwFM & 0.3617$\pm$0.00026 & 0.7335$\pm$0.00071 & 1.705\% \\
    \midrule
    DeepFM & 0.3532$\pm$0.00050 & 0.7370$\pm$0.00042 & 2.191\% \\
    AutoInt+ & 0.3604$\pm$0.00045 & 0.7313$\pm$0.00076 & 1.400\% \\
    PNN & 0.3523$\pm$0.00028 & 0.7364$\pm$0.00025 & 2.108\% \\
    \midrule
    DIN & 0.3510$\pm$0.00014 & 0.7394$\pm$0.00063 & 2.524\% \\
    DIEN & 0.3485$\pm$0.00032 & 0.7411$\pm$0.00058 & 2.759\% \\
    DSIN & 0.3477$\pm$0.00011 & 0.7442$\pm$0.00017 & 3.189\% \\
    \midrule
    GIN & 0.3508$\pm$0.00046 & 0.7328$\pm$0.00072 & 1.608\% \\
    Fi-GNN & 0.3516$\pm$0.00037 & 0.7339$\pm$0.00025 & 1.761\% \\
    DG-ENN & \underline{0.3416}$\pm$0.00053 & \underline{0.7489}$\pm$0.00064 & 3.841\% \\
    \midrule
    HIEN (ours) & \textbf{0.3402$\pm$0.00027} & \textbf{0.7526$\pm$0.00013} & \textbf{4.354\%}\\
    \bottomrule
    \end{tabular}
    \label{tab:exp_production}
\end{table}

The experiment results of comparison between existing CTR prediction models and our proposed HIEN are shown in Tab.~\ref{tab:exp_public} and~\ref{tab:exp_production} on three datasets, respectively. We have the following observations:
\begin{itemize}
    \item HIEN consistently achieves the best performance, and it outperforms the best baseline (DG-ENN) by 0.86\%, 0.76\% and 0.49\% in terms of AUC (4.81\%, 2.08\% and 0.41\% in terms of LogLoss) on Alibaba, Tmall and Tencent datasets, respectively. Possible reasons for the great improvement of HIEN over state-of-the-art methods may be the attribute tree aggregation for considering structure information and relations among attributes as well as the intent modeling. We will further validate its effectiveness in later experiments.
    \item LR is the worst method among all baselines since it only models a shallow linear combination of features. FM and FwFM perform better than LR, which proves that the second-order feature interactions are effective in CTR prediction. DeepFM, AutoInt+, and PNN perform better than shallow models due to high-order feature interaction modeling through DNN. User interest modeling is another practical way to improve performance, such as DIN, DIEN, and DSIN.
    \item GIN and Fi-GNN apply graph convolution to model feature interactions, while DG-ENN considers the complex relations between users and items. However, these models neglect structure information of attributes and user/item intents for different attributes. Some of these latest graph models are more computationally complex than our HIEN, yet HIEN still delivers better CTR prediction performance.
\end{itemize}

It is worth noting that, when considering a large user base, an improvement of AUC at 0.1\% level is generally considered as practically significant for industrial CTR prediction task~\cite{cheng2016wide}. This fact has been recognized by several existing studies from Google~\cite{cheng2016wide} and Huawei~\cite{guo2017deepfm}. As reported in~\cite{cheng2016wide}, compared with LR, Wide \& Deep improves AUC by 0.275\% (ofﬂine) and the improvement of online CTR is 3.9\%. Even several percents lift in CTR brings extra millions of dollars each year. 

Moreover, we conduct a t-test between our proposed HIEN and the baselines. The p-value of HIEN against all baselines under LogLoss or AUC is less than 1e-6 on three datasets, which indicates that our improvement over existing models is signiﬁcant. 
% In addition, our model makes better CTR prediction but with less computing power (15\% less training time) compared to the best baseline model DG-ENN.

\subsection{Study of HIEN (RQ2)}
\label{sec:hien_study}
\subsubsection{Impact of Aggregators} In order to explore the impact of different aggregators in HIEN, as formulated in Eqn.~\eqref{eqn:gcn}-\eqref{eqn:cp-agg}, we compare the performance of our proposed model with different aggregators. The experiment results are shown in Tab.~\ref{tab:exp_aggregator}. We can find that CP-Agg achieves the best performance due to modeling two kind of feature interactions. Besides, NGCF performs better than GCN aggregator, since it considers feature interactions between central node and neighbor nodes. Moreover, the effect of LightGCN is not as good as expected. A possible reason is that LightGCN removes central node representation. However, the graph structure is relatively simple in online advertising scenario compared to traditional recommendation system, which leads to the fact that only using neighbor nodes to represent central nodes is inadequate.

\begin{table}[ht]
    \centering
    \caption{Effect of different aggregators in HIEN on three datasets.}
    \setlength{\tabcolsep}{1.5mm}{
    \begin{tabular}{c|cc|cc|cc}
    \toprule
    \multirow{2}{*}{Aggregator} & \multicolumn{2}{c|}{Alibaba} & \multicolumn{2}{c|}{Tmall} & \multicolumn{2}{c}{Tencent}\\
     & Loss & AUC & Loss & AUC & Loss & AUC\\
    \midrule
    GCN & 0.2175 & 0.6781 & 0.1623 & 0.9452 & 0.3415 & 0.7506\\
    NGCF & 0.2167 & 0.6789 & 0.1612 & 0.9461 & 0.3408 & 0.7520\\
    LightGCN & 0.2172 & 0.6784 & 0.1620 & 0.9457 & 0.3412 & 0.7514\\
    CP-Agg & \textbf{0.2156} & \textbf{0.6798} & \textbf{0.1601} & \textbf{0.9467} & \textbf{0.3402} & \textbf{0.7526}\\
    \bottomrule
    \end{tabular}
    }
    \label{tab:exp_aggregator}
\end{table}

\subsubsection{Ablation study of HIEN} We conduct experiments on three datasets to validate the effectiveness of different components and how these components contribute to the overall results, including attribute tree aggregation and intent modeling. Correspondingly, we design a series of ablation studies for HIEN. Four variants are considered to simplify HIEN in different ways: 1) removing user attribute tree aggregation, 2) removing item attribute tree aggregation, 3) removing user intent modeling, 4) removing item intent modeling. Tab.~\ref{tab:exp_ablation1}, Tab.~\ref{tab:exp_ablation3}, and Tab.~\ref{tab:exp_ablation2} show the results of LogLoss and AUC for these variants, and their relative performance drop compared with the baseline, i.e., the original HIEN.

We observe that the second variant (w/o item\_agg) performs worst among all variants, which indicates that the item attribute tree involves rich structure information compared to the user attribute tree for improving performance. Besides, the third variant (w/o user\_intent) is also greatly affected, which suggests that modeling user intents for different item attributes is effective in our prediction tasks. Other variants are also affected to varying degrees.

\begin{table}[ht]
    \centering
    \caption{Effect of different components in HIEN on public Alibaba dataset.}
    \setlength{\tabcolsep}{3mm}{
    \begin{tabular}{c|c|c}
    \toprule
    Variants & Loss & AUC\\
    \midrule
    w/o user\_agg & 0.2175 (+0.0019) & 0.6796 (-0.0002)\\
    w/o item\_agg & 0.2248 (+0.0092) & 0.6783 (-0.0015)\\
    w/o user\_intent & 0.2236 (+0.0080) & 0.6786 (-0.0012)\\
    w/o item\_intent & 0.2204 (+0.0048) & 0.6791 (-0.0007)\\
    \midrule
    HIEN (original) & \textbf{0.2156} & \textbf{0.6798}\\
    \bottomrule
    \end{tabular}}
    \label{tab:exp_ablation1}
\end{table}

\begin{table}[ht]
    \centering
    \caption{Effect of different components in HIEN on public Tmall dataset.}
    \setlength{\tabcolsep}{3mm}{
    \begin{tabular}{c|c|c}
    \toprule
    Variants & Loss & AUC\\
    \midrule
    w/o user\_agg & 0.1607 (+0.0006) & 0.9459 (-0.0008)\\
    w/o item\_agg & 0.1630 (+0.0029) & 0.9428 (-0.0039)\\
    w/o user\_intent & 0.1625 (+0.0024) & 0.9432 (-0.0035)\\
    w/o item\_intent & 0.1614 (+0.0013) & 0.9450 (-0.0017)\\
    \midrule
    HIEN (original) & \textbf{0.1601} & \textbf{0.9467}\\
    \bottomrule
    \end{tabular}}
    \label{tab:exp_ablation3}
\end{table}

\begin{table}[ht]
    \centering
    \caption{Effect of different components in HIEN on production dataset.}
    \setlength{\tabcolsep}{3mm}{
    \begin{tabular}{c|c|c}
    \toprule
    Variants & Loss & AUC\\
    \midrule
    w/o user\_agg & 0.3410 (+0.0008) & 0.7521 (-0.0005)\\
    w/o item\_agg & 0.3419 (+0.0017) & 0.7496 (-0.0030)\\
    w/o user\_intent & 0.3418 (+0.0016) & 0.7511 (-0.0015)\\
    w/o item\_intent & 0.3412 (+0.0010) & 0.7517 (-0.0009)\\
    \midrule
    HIEN (original) & \textbf{0.3402} & \textbf{0.7526}\\
    \bottomrule
    \end{tabular}}
    \label{tab:exp_ablation2}
\end{table}

\subsubsection{Effect of layer number} 
\begin{figure*}[!htbp]
	\begin{subfigure}[b]{0.26\linewidth}
		\centering
		\includegraphics[width=\linewidth]{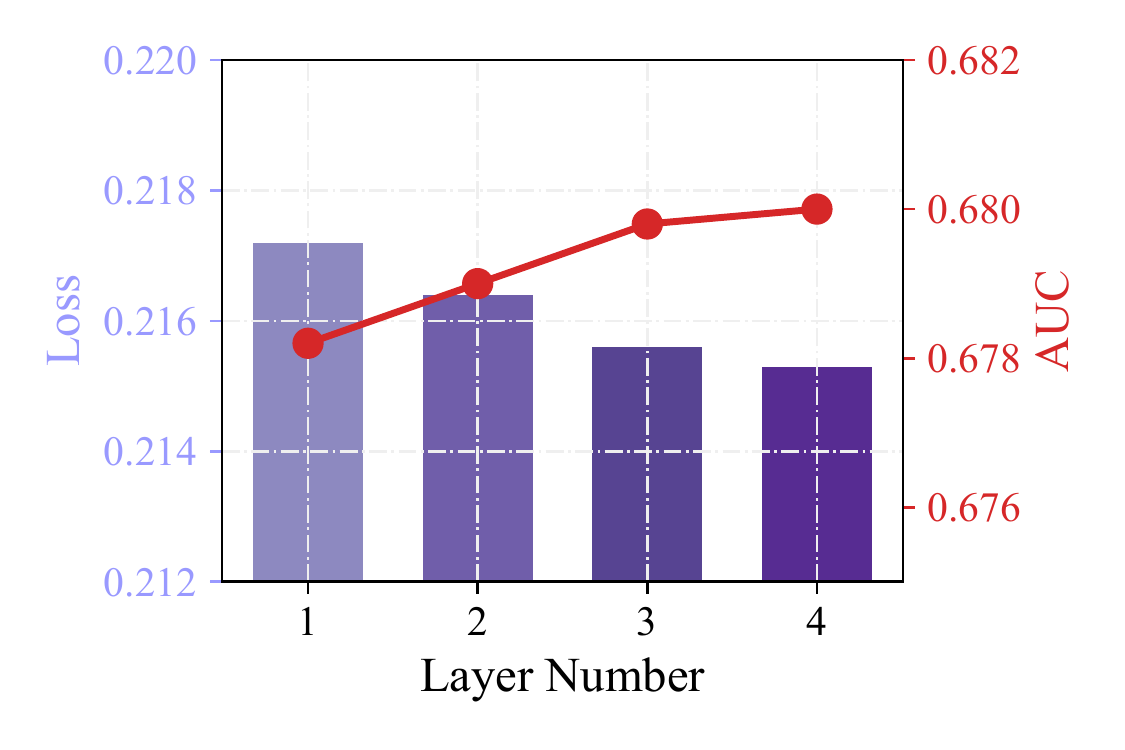}
		\caption{Alibaba}
		\label{fig:ln1}
	\end{subfigure}%
	\begin{subfigure}[b]{0.26\linewidth}
		\centering
		\includegraphics[width=\linewidth]{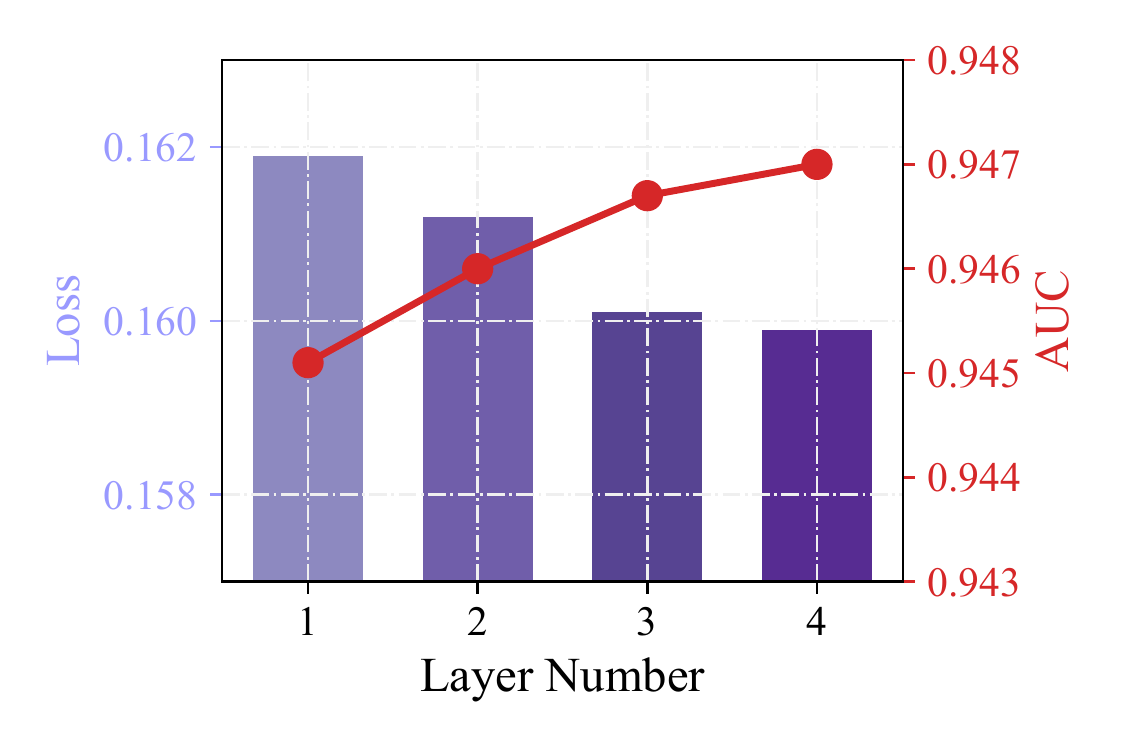}
		\caption{Tmall}
		\label{fig:ln1}
	\end{subfigure}%
	\begin{subfigure}[b]{0.26\linewidth}
		\centering
		\includegraphics[width=\linewidth]{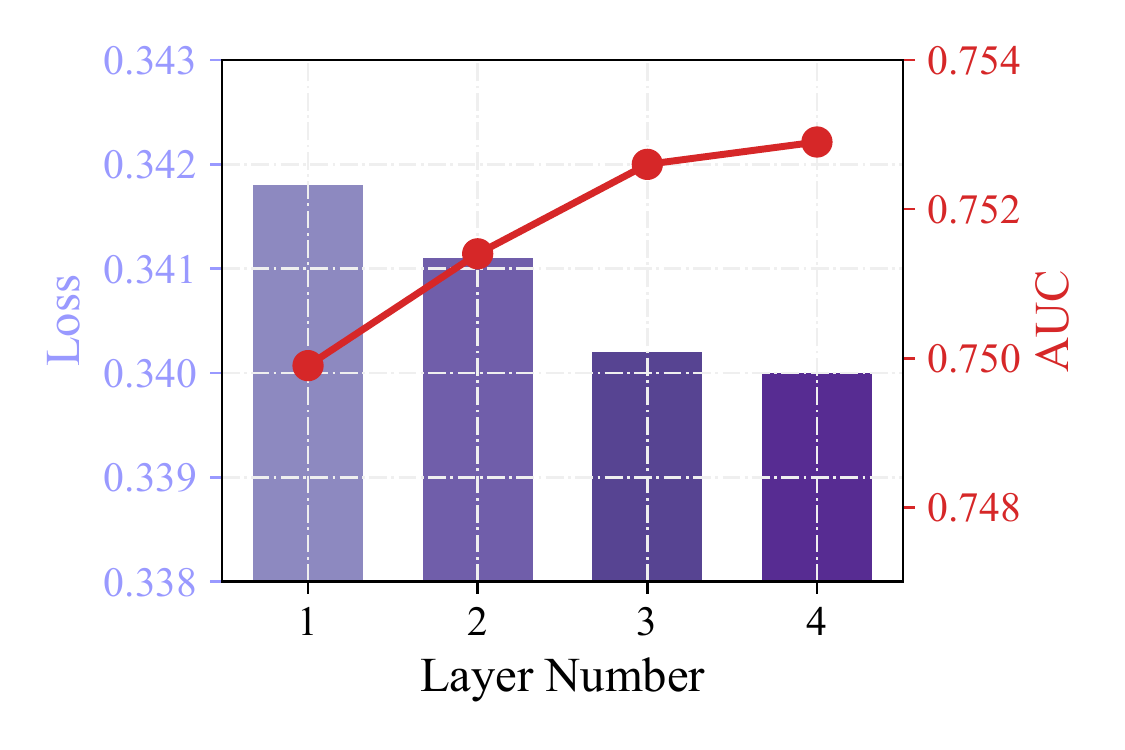}
		\caption{Tencent}
		\label{fig:ln2}
	\end{subfigure}%
	\caption{Effect of embedding propagation layer numbers $L$ on three dataset.}
	\label{exp:layer_number}
\end{figure*}

We vary the depth of HIEN (layer number) to investigate the efficiency of usage of multiple embedding propagation layers when learning user and item representations. In particular, the layer number is searched in the range of $\{1, 2, 3, 4\}$; we use HIEN-1 to indicate the model with one layer, and similar notations for others. We summarize the results in Fig.~\ref{exp:layer_number}, and have the following observations: a) we can observe that increasing the layer number is capable of boosting the performance substantially. HIEN-2 and HIEN-3 achieve consistent improvement over HIEN-1 across all the board, since the effective modeling of high-order relations between users and items; b) When stacking one more layer over HIEN-3, we find that the improvement of HIEN-4 is marginal. It indicates that considering third-order relations among graph nodes could be sufficient to learn user and item embeddings.

\subsection{HIEN as an Input Module (RQ3)}
Through our proposed HIEN, the attribute embedding and user/item embedding are refined, which can be used as a general input module in other CTR models. In order to explore the compatibility of our proposed HIEN, we integrate FwFM, DeepFM, DSIN, and DG-ENN with the refined embedding of HIEN as input. The experiment results are shown in Tab.~\ref{tab:exp_input}. We can observe that the performance of these methods can be further boosted with our proposed HIEN. For example, DSIN outperforms its original method by 1.10\%, 0.72\%, and 1.89\% in terms of AUC (7.11\%, 4.81\%, and 3.30\% in terms of LogLoss) on three datasets.
It validates the compatibility of our refined embedding by demonstrating its effectiveness in working with various popular CTR models. The refined embedding involves more information with structures and relations between attributes and users/items.

\begin{table}[ht]
    \centering
    \caption{Compatibility of refined embedding by HIEN on three datasets. $+$ denotes the new model integrated with HIEN as input module.}
    \setlength{\tabcolsep}{1.5mm}{
    \begin{tabular}{l|cc|cc|cc}
    \toprule
    \multirow{2}{*}{Model} & \multicolumn{2}{c|}{Alibaba} & \multicolumn{2}{c|}{Tmall} & \multicolumn{2}{c}{Tencent}\\
    & Loss & AUC & Loss & AUC & Loss & AUC\\
    \midrule
    FwFM & 0.2526 & 0.6427 & 0.1791 & 0.9075 & 0.3617 & 0.7335\\
    FwFM+& \textbf{0.2497} & \textbf{0.6482} & \textbf{0.1765} & \textbf{0.9114} & \textbf{0.3584} & \textbf{0.7347}\\
    \midrule
    DeepFM & 0.2447 & 0.6594 & 0.1772 & 0.9184 & 0.3532 & 0.7370\\
    DeepFM+& \textbf{0.2184} & \textbf{0.6771} & \textbf{0.1752} & \textbf{0.9285} & \textbf{0.3402} & \textbf{0.7526}\\
    \midrule
    DSIN & 0.2308 & 0.6708 & 0.1684 & 0.9313 & 0.3477 & 0.7442\\
    DSIN+& \textbf{0.2144} & \textbf{0.6782} & \textbf{0.1603} & \textbf{0.9380} & \textbf{0.3362} & \textbf{0.7583} \\
    \midrule
    DG-ENN & 0.2265 & 0.6740 & 0.1635 & 0.9396 & 0.3416 & 0.7489\\
    DG-ENN+& \textbf{0.2107} & \textbf{0.6816} & \textbf{0.1590} & \textbf{0.9485} & \textbf{0.3328} & \textbf{0.7596}\\
    \bottomrule
    \end{tabular}}
    \label{tab:exp_input}
\end{table}

\subsection{Visualization of Intent Modeling (RQ4)}
\begin{figure}[h]
	\centering
	\includegraphics[width=\linewidth,angle=0]{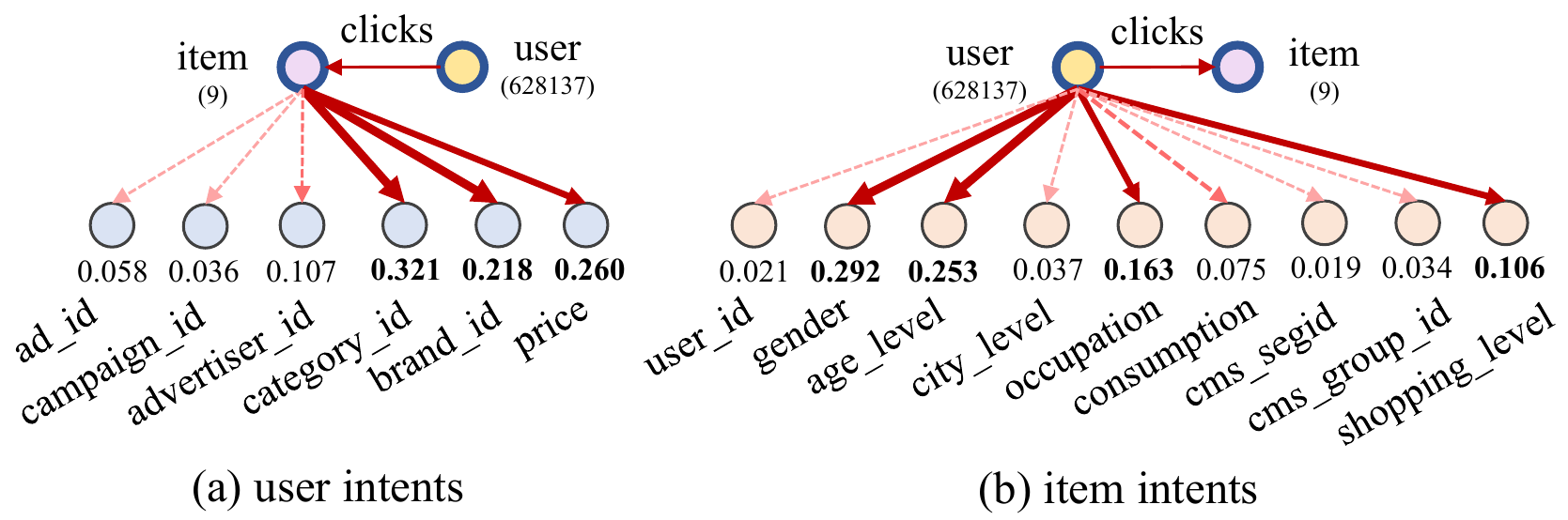}
	\caption{Explanations of intent modeling and real case on the public Alibaba dataset.}
	\label{fig:intents}
\end{figure}
In this section, we present a real case of intent modeling on the public Alibaba dataset. As shown in Fig.~\ref{fig:intents}, we observe that some item attributes indeed play important roles in user intents when a user (628137) clicks an item (9), including \emph{category\_id}, \emph{brand\_id}, and \emph{price}. Besides, it can be found that some user attributes get high weights in item intents, such as \emph{gender}, \emph{age\_level}, \emph{occupation}, and \emph{shopping\_level}. These observations indicate that such attributes are common factors pertinent to user behaviors.

\section{Related Works}
% In this section, we discuss three research areas related to our works, including feature interaction modeling, user interest modeling, and graph neural network for recommendation.
\subsection{Feature Interaction Modeling}
Click-through Rate (CTR) Prediction is one of the fundamental tasks in the online advertising and recommendation system, which aims at predicting the probability that a user clicks an item or ad. Pioneer works are proposed mainly based on Logistic Regression (LR)~\cite{richardson2007predicting,chapelle2014simple,mcmahan2013ad}, polynomial~\cite{chang2010training}, collaborative filtering~\cite{shen2012personalized}, Bayesian models~\cite{graepel2010web}, etc. 
In order to explicitly model the feature interactions, many factorization machine based methods are proposed for high-dimensional data, such as Factorization Machine (FM)~\cite{rendle2010factorization}, Field-aware Factorization Machine (FFM)~\cite{juan2016field}, Field-weighted Factorization Machine (FwFM)~\cite{pan2018field,pan2019predicting}, and Field-matrixed Factorization Machine (FmFM)~\cite{sun2021fm}. Besides, there are some works that aim at learning weight for different feature interactions, including Attentional Factorization Machines (AFM)~\cite{xiao2017attentional}, Dual-attentional Factorization Machines (DFM)~\cite{liu2020dual}, Dual Input-aware Factorization Machines (DIFM)~\cite{lu2020dual}.

Since the number of samples and the dimension of features have become increasingly larger, many deep learning based models have been proposed to learn high-order feature interactions, such as Wide\&Deep~\cite{cheng2016wide}, Deep Crossing~\cite{shan2016deep}, YouTube Recommendation~\cite{covington2016deep}, PNN~\cite{qu2016product}, Deep\&Cross~\cite{wang2017deep}, AutoInt+~\cite{DBLP:conf/cikm/SongS0DX0T19}. Some studies combine FM with DNN, such as DeepFM~\cite{guo2017deepfm}, xDeepFM~\cite{lian2018xdeepfm}, NFM~\cite{he2017neural}, and DeepLight~\cite{deng2021deeplight}. Overall, these models follows Embedding\&Multi-Layer Perceptron (MLP) paradigm. However, they regard item attributes as ID features, while neglecting the relations and dependencies between attributes of an item.

\subsection{User Interest Modeling}
Traditional user interest methods take a straightforward way to represent each behavior with an embedding vector, and then do a sum or mean pooling over all these embedding vectors to generate one embedding~\cite{covington2016deep}.
Deep Interest Network (DIN)~\cite{zhou2018deep} first considers the effect of different target items, and assigns attentive weights for user behaviors. It could captures the user interests on the different target items. Deep Interest Evolution Network (DIEN)~\cite{zhou2019deep} uses GRU encoder to capture the dependencies of user behaviors, followed by another GRU with an attentional update gate to depict interest evolution. Deep Session Interest Network (DSIN)~\cite{feng2019deep} leverages users' multiple historical sessions in their behavior sequences, which is based on Transformer and Bi-LSTM. There are also some works that further consider long-term historical behavior sequences, such as Multi-channel user Interest Memory Network (MIMN)~\cite{pi2019practice}, Hierarchical Periodic Memory Network (HPMN)~\cite{ren2019lifelong}, and Search-based Interest Model (SIM)~\cite{pi2020search}.
% Besides, Kalman Filtering Attention (KFAtt)~\cite{liu2020kalman} addresses the limitations of limited attention fields within a single user’s behaviors and the bias towards frequent behaviors based on Kalman Filtering. User Behavior Retrieval for CTR prediction (UBR4CTR)~\cite{DBLP:conf/sigir/Qin0WJF020} retrieves the most relevant and appropriate user behaviors from the entire user history sequence instead of simply using the most recent ones. Multi-Interactive Attention Network (MIAN)~\cite{DBLP:conf/wsdm/ZhangQCLLZMC21} comprehensively extract the latent relationship between the target item and all kinds of fine-grained features, including behavior, profile, and context.

However, these user behavior models focus on mining interests through user-item interactions, while ignoring user intents and item intents for different attributes.

\subsection{Graph Neural Network for Recommendation}
Graph Neural Network is widely used in recommender systems in recent years and achieves significant success. Graph Intention Network (GIN)~\cite{DBLP:conf/sigir/LiCWRZZ19} adopts multi-layered graph diffusion to enrich user behaviors, which can solve the behavior sparsity problem. By introducing the co-occurrence relationship of commodities to explore the potential preferences, the weak generalization problem is also alleviated. Feature Interaction Graph Neural Networks (Fi-GNN)~\cite{DBLP:conf/cikm/LiCWZW19} represents the multi-field features in a graph structure, where each node corresponds to a feature field and different fields can interact through edges. The task of modeling feature interactions can be thus converted to modeling node interactions on the corresponding graph. Dual Graph enhanced Embedding Neural Network (DG-ENN)~\cite{DBLP:conf/kdd/GuoSTGZLTH21} exploits the strengths of graph representation with two carefully designed learning strategies (divide-and-conquer, curriculum-learning-inspired organized learning) to refine the embeddings. There are also some GNN based models in recommendation systems, such as LightGCN~\cite{DBLP:conf/sigir/0001DWLZ020}, Neural Graph Collaborative Filtering (NGCF)~\cite{DBLP:conf/sigir/Wang0WFC19}. Besides, to utilize knowledge graph (KG) and make full use of other information beyond user-item interactions, Knowledge Graph Attention Network (KGAT)~\cite{DBLP:conf/kdd/Wang00LC19} is proposed, which explicitly models the high-order connectivities in KG in an end-to-end fashion. Knowledge Graph-based Intent Network (KGIN)~\cite{DBLP:conf/www/WangHWYL0C21} explores intents behind a user-item interaction, which are modeled as an attentive combination of KG relations.

\section{Conclusion}
In this paper, we focus on the structural feature embedding learning and intent modeling in the CTR prediction scenario of online advertising systems. In the CTR prediction model, feature interaction modeling and user interest modeling methods are two popular domains. Currently, these CTR prediction models still suffer from two limitations. First, traditional methods regard item attributes as ID features, while neglecting structure information and relation dependencies among attributes. Second, when mining user interests from user-item interactions, current models fail to reveal user intents and item intents for different attributes, which lacks interpretability.
To improve the expressive ability and effectiveness of the CTR model, we propose a novel Hierarchical Intention Embedding Network (HIEN), considering dependencies of attributes based on bottom-up tree aggregation. In addition, HIEN can capture user intents for different item attributes as well as item intents based on the hierarchical attention mechanism proposed in this paper.
Extensive experiments are conducted on both public and production production datasets, and the results show that the proposed model significantly outperforms the state-of-the-art models. Moreover, our structure embedding learning technique can serve as an input module for existing state-of-the-art CTR prediction methods to boost their performance.

% \begin{acks}

% \end{acks}

\newpage
\bibliographystyle{ACM-Reference-Format}
\balance
\bibliography{ref}

\end{document}